\documentclass[11pt,twoside]{article}
\usepackage{asp2010}

\resetcounters

\begin{document}

\title{Coronal Polarization}
\author{N.-E. Raouafi}
\affil{Johns Hopkins University Applied Physics Laboratory, Laurel, MD, USA}

\begin{abstract}
We present an overview of the physical mechanisms responsible for the coronal polarization at different wavelength regimes. We also review different techniques using coronal polarization to determine various quantities necessary for understanding the thermodynamic properties of the solar coronal plasma. This includes the coronal magnetic field, electronic densities, temperatures, velocities, etc. The future needs to acquire better information on the solar corona using polarization will be outlined.
\end{abstract}

\vspace{-1cm}
\section{Introduction}

Our understanding of coronal manifestations, such as the coronal heating, the acceleration of the solar wind, solar magnetic activity, energetic particles, etc., faces a major handicap caused mainly by the lack of measurements of key parameters such as densities, temperatures, magnetic fields, and velocities. Understanding the origin and evolution of coronal phenomena is not only important from a scientific point of view, but also because their effects extend from the Sun till the outer edges of the solar system throughout the interplanetary medium including the Earth environment.

Other considerations also make coronal studies a priority for characterizing this environment. The National Aeronautics and Space Administration (NASA) and the European Space Agency (ESA) have plans to launch spacecrafts to probe the near-Sun environment. Missions such as the Solar Probe Plus (SPP) and the Solar Orbiter (SO) will approach the Sun closer than ever ($\sim9.5$ R$_{\sun}$ for SPP and $\sim0.28$ AU for SO). The success of these missions and the quality of the measurements depends greatly on the design of spacecrafts that resist the effects of the harsh environment they will fly through. However, in order to achieve such a successful design, there is a need to know the inner corona better than ever.

The polarization of coronal emissions at different wavelength regimes is an excellent tool to probe key plasma parameters such as densities and magnetic fields. This paper presents a brief overview of the different techniques using coronal polarization to diagnose the plasma properties. Thomson scattering and measurements of the magnetic fields are outlined sections 2 and 3, respectively. The prospects of coronal polarization will be discussed in section 4. 

\section{Polarization of the K-corona: Thomson scattering}

The white-light emission in the solar corona arises from the scattering of photospheric light through three main mechanisms: the K-corona is the result of the scattering of radiation by free coronal electrons (i.e., Thomson scattering), scattering by dust particles form the F-corona, and atomic spectral lines contribute a relatively small fraction that constitutes the E-corona. Since the corona is optically thin, the observer receives contribution from a wide interval along the line of sight (LOS).

Despite the lack of observations, the theory of coronal Thomson scattering was studied  as early as 1879 by Schuster and refined later by \cite{1930ZA......1..209M}. The expression linking the coronal polarized brightness ($pB_K$) to the electron density is as follow:\\
\centerline{$\displaystyle pB_K \propto \int_{LOS} N_e(r)\;G(\rho,s) ds,$} 
the full expression and details of the different parameters (i.e., $G$, $r$, $\rho$, and $s$) are given by \cite{2002A&A...393..295Q}. The polarization measurements are usually used to separate the K- and F-coronas. This is based on the assumption that polarized brightness, $pB_K$, does not depend on the brightness of the F-corona, which is only valid below 5~R$_{\sun}$. At greater altitudes, the polarization of the F-corona has to be accounted for \citep[see][]{1985ASSL..119...63K,1992A&A...261..329M,2001ApJ...548.1081H}.

\vspace{-0.25cm}
\begin{figure}[!h]
   \begin{center}
\parbox{8.7cm}{\includegraphics[width=8.7cm]{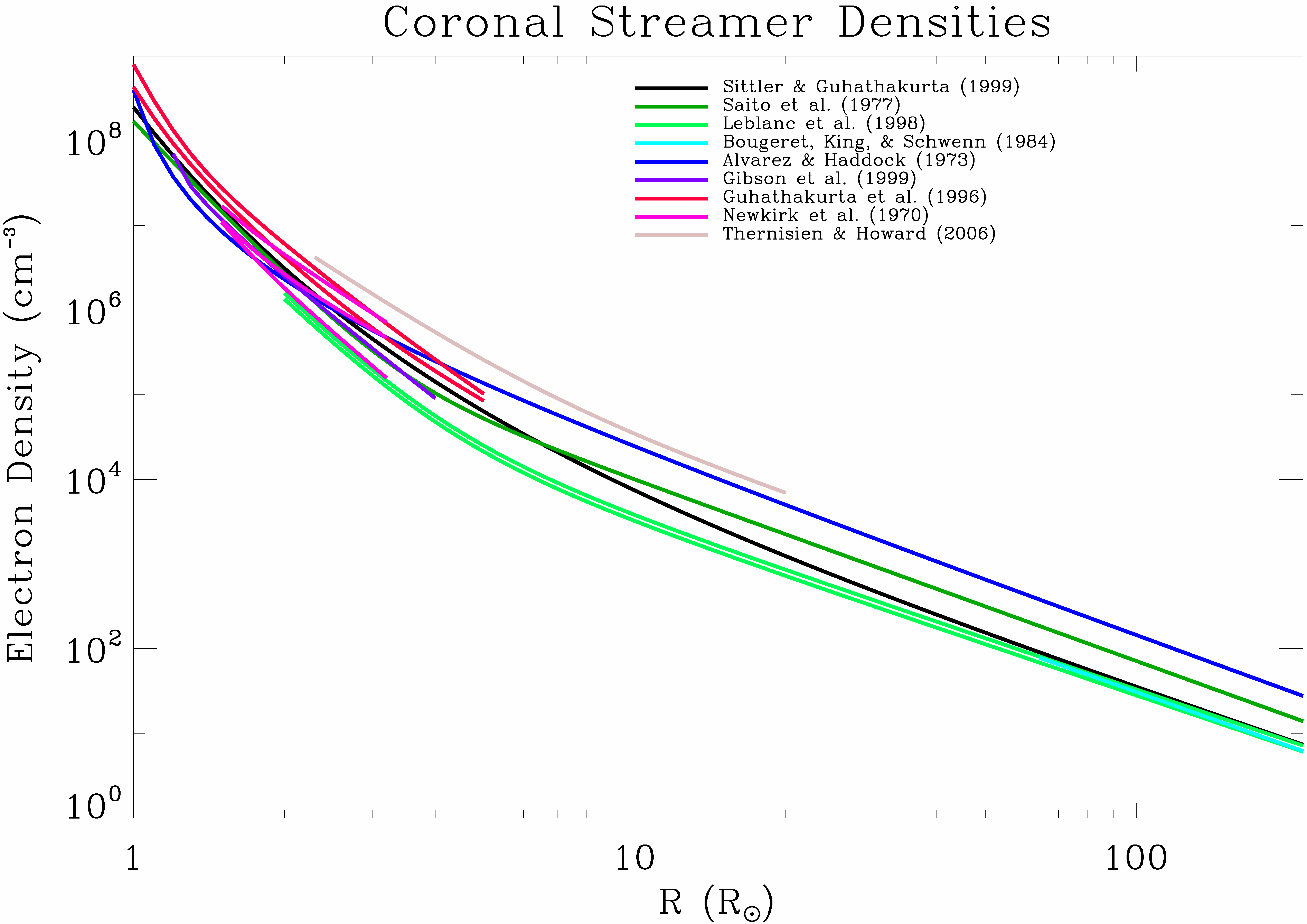}}
\parbox{5.2cm}{\caption{Streamer electron densities from different measurements and models. Some of the data are extrapolated to 1 AU. \label{Streamer_coronal_densities_spw6} }}
   \end{center}
\vspace{-0.8cm}
\end{figure}

\cite{1950BAN....11..135V} presented an inversion method for deriving the coronal electron density from white-light polarization observations. The main assumption is power-law representations for the polarized radiance and the electron density as a function of the heliodistance. Although this method is simple, it is so successful that is still applied to every polarization brightness measurements. Figure~\ref{Streamer_coronal_densities_spw6} is an illustration of density measurements and models within coronal streamers. Most of the measurements between 1~R$_{\sun}$ and a few solar radii are based on the polarization brightness of coronal white light.

Recent inversion models are improved significantly to account for the quality and quantity of the available data. For instance, the complementarity of the coronal total and polarized brightness is taken into consideration in most models \citep[see, e.g.,][]{2002A&A...393..295Q}. Some other approaches include more sophisticated models of coronal structure, such as the slab model of streamers \citep{1996ApJ...458..817G,1997ASPC..125..230V,2006ApJ...642..523T}.

\section{Magnetic Field Measurements}

The coronal magnetic fields can affect the polarization of the emitted radiation through different physical mechanisms. Conversely, the polarization of coronal emissions can provide an excellent and may be unique (in the lack of in situ measurements) diagnostic tool of the coronal magnetic fields.

\subsection{Zeeman Effect}

Because of the proportionality of the Zeeman splitting to $\lambda^2$, spectral line with longer wavelength (i.e., infra-red [IR]) are better suited for studying coronal magnetic fields.

\begin{figure*}[!h]
\begin{center}
\parbox{9.3cm}{\includegraphics[width=9.9cm]{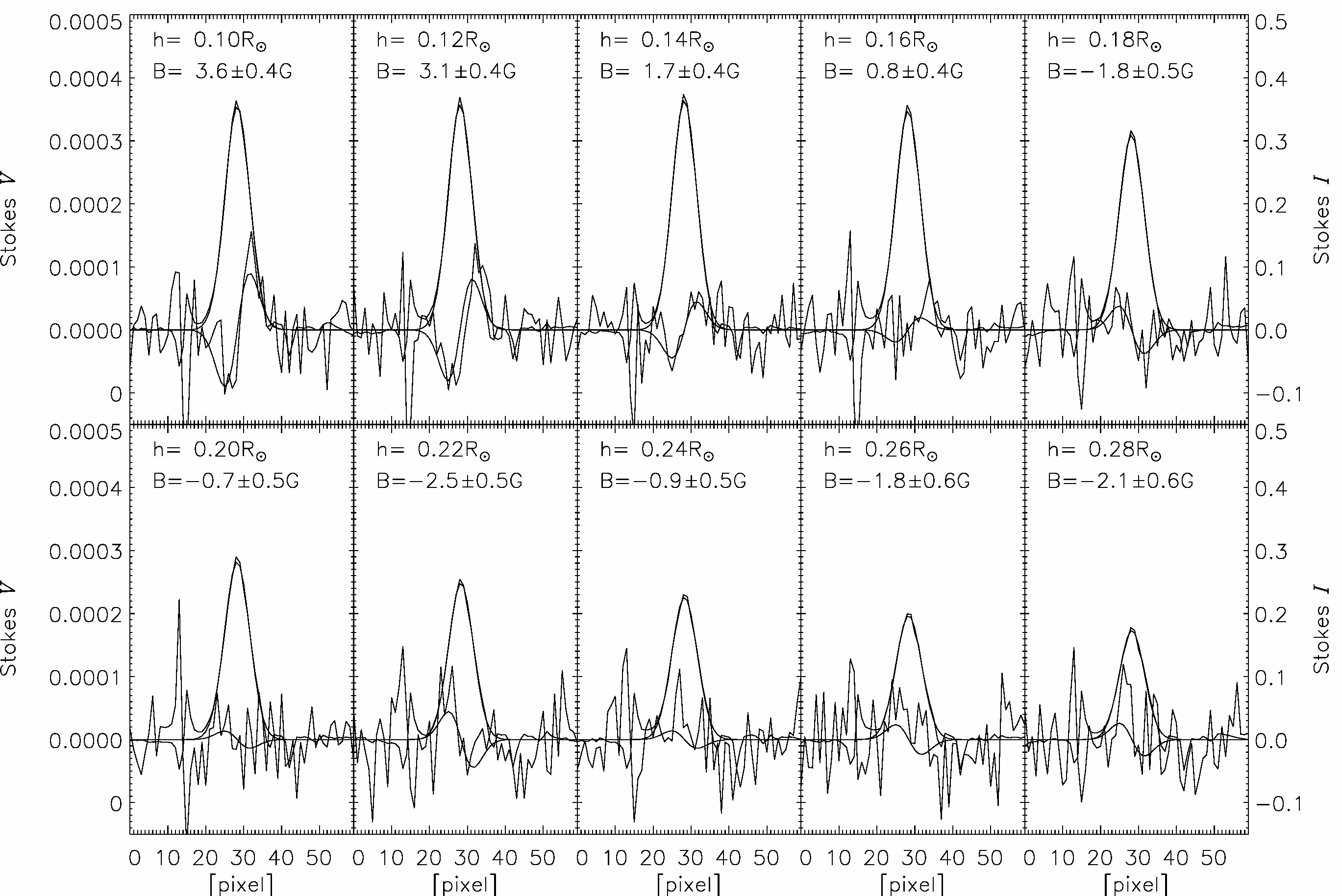}}
\parbox{4.7cm}{\caption{Stokes I and V profiles (observed and fitted) of the \ion{Fe}{xiii} 1074.7 nm line from different distances above NOAA AR 581 on April 06, 2004. The magnetic field strength are shown for each data set. From \cite{2004ApJ...613L.177L}. \label{vmap}}}
\end{center}
\vspace{-0.75cm}
\end{figure*}

The coronal Zeeman effect is mainly restricted to active regions (ARs) at the limb and to IR lines. \cite{1969PhDT.........3H} used the \ion{Fe}{xiv} 530.3 nm green line to measure the LOS component of magnetic fields above ARs. Despite instrumental limitations, he obtained measurements of the LOS-field strength of the order of $13\pm20$~Gauss. \cite{1995itsa.conf...89K} used the IR \ion{Fe}{xiii} 1074.7 nm line to set an upper limit on the LOS-field strength above ARs ($\sim40$ Gauss). Recently, more accurate measurements have been achieved by \cite{2000ApJ...541L..83L,2004ApJ...613L.177L} using the same IR line to measure fields at heights ranging between 1.1 and 1.28 R$_{\sun}$ from Sun center. Figure~\ref{vmap} illustrates the measured Stokes I and V of these observations and the corresponding magnetic field values.

Observing the Zeeman effect in the corona has to overcome not only the difficulties in detecting the small signal due to the weak magnetic field, but mainly the hurdle of limited development of dedicated instruments. Although \cite{2000ApJ...541L..83L,2004ApJ...613L.177L} have proved the possibility of observing the Zeeman effect in the corona, they had to deal with excessively long time exposures and a lack of spatial resolution due mainly to the limited aperture of the telescope.

\subsection{Hanle Effect}

The Hanle effect is the modification of the linear polarization of a spectral line by a local magnetic field \citep{1924ZPhy...30...93H}. Theoretically, direct determination of the magnetic field in the solar corona could be achieved through linear polarization of spectral lines with suitable sensitivity to the Hanle effect.

The Hanle effect in selected spectral lines yields a powerful diagnostic tool for magnetic fields ranging typically from one milligauss to several 100 Gauss (depending strongly on the chosen line and the strength and direction of the magnetic field). Unlike the Zeeman effect, the depolarization of spectral lines by turbulent magnetic fields can be detected in the Hanle regime allowing the determination of the strength of the field in mixed-polarity regions \citep[e.g.,][]{1982SoPh...80..209S,2004Natur.430..326T}.

The Hanle effect diagnostic of magnetic fields has been successful in solar prominences \citep{1977A&A....54..811L,1977A&A....59..223S,1980A&A....87..109B,1982SoPh...79..291L,1985SoPh...96..277Q,2002Natur.415..403T,2002ApJ...575..529L}.

\subsubsection{Polarization of Coronal Forbidden Lines}

\cite{1965AnAp...28..877C} has shown that the direction of polarization of some forbidden lines is expected to be either parallel or perpendicular to the local magnetic field projected onto the plane of the sky. This provides a fruitful approach to study the orientation (direction) of the coronal magnetic field. No information on the field strength can, however, be obtained from such diagnostics.

The polarization of forbidden lines such as \ion{Fe}{xiv} 530.3~nm and \ion{Fe}{xiii} 1074.7~nm have been studied for more than three decades during solar eclipses and coronagraphic observations \citep[][etc.]{1974psns.coll..254Q,1977ROLun..12..109Q,1987A&A...178..263A}. Striking evidence of a predominant radial orientation of the polarization is found everywhere independently of the phase of the solar cycle, which depicts the direction of the coronal field projected on the plane of the sky.

More recently, \cite{2001ApJ...558..852H} compared intensity and polarization maps with better resolution of the \ion{Fe}{xiii} 1074.7~nm line from 1980. They found evidence for two magnetic components in the corona: a non-radial field associated with the large-scale structures known as streamers (with loop-like structures at their base) and a more pervasive radial magnetic field, which corresponds to the open coronal magnetic field. They also noticed that using pB images alone is not a reliable approach for the inference of the direction of the coronal magnetic field.

\subsubsection{Polarization of FUV and EUV Coronal Lines}

Several lines in the FUV (far UV) wavelength range present suitable sensitivity to determine the coronal magnetic field via the Hanle effect. The coronal Hanle effect in the FUV and EUV (extreme UV) wavelength ranges is largely unexplored despite the high potential of this diagnostic.

\begin{figure}[!th]
\begin{center}
\includegraphics[width=.35\textwidth]{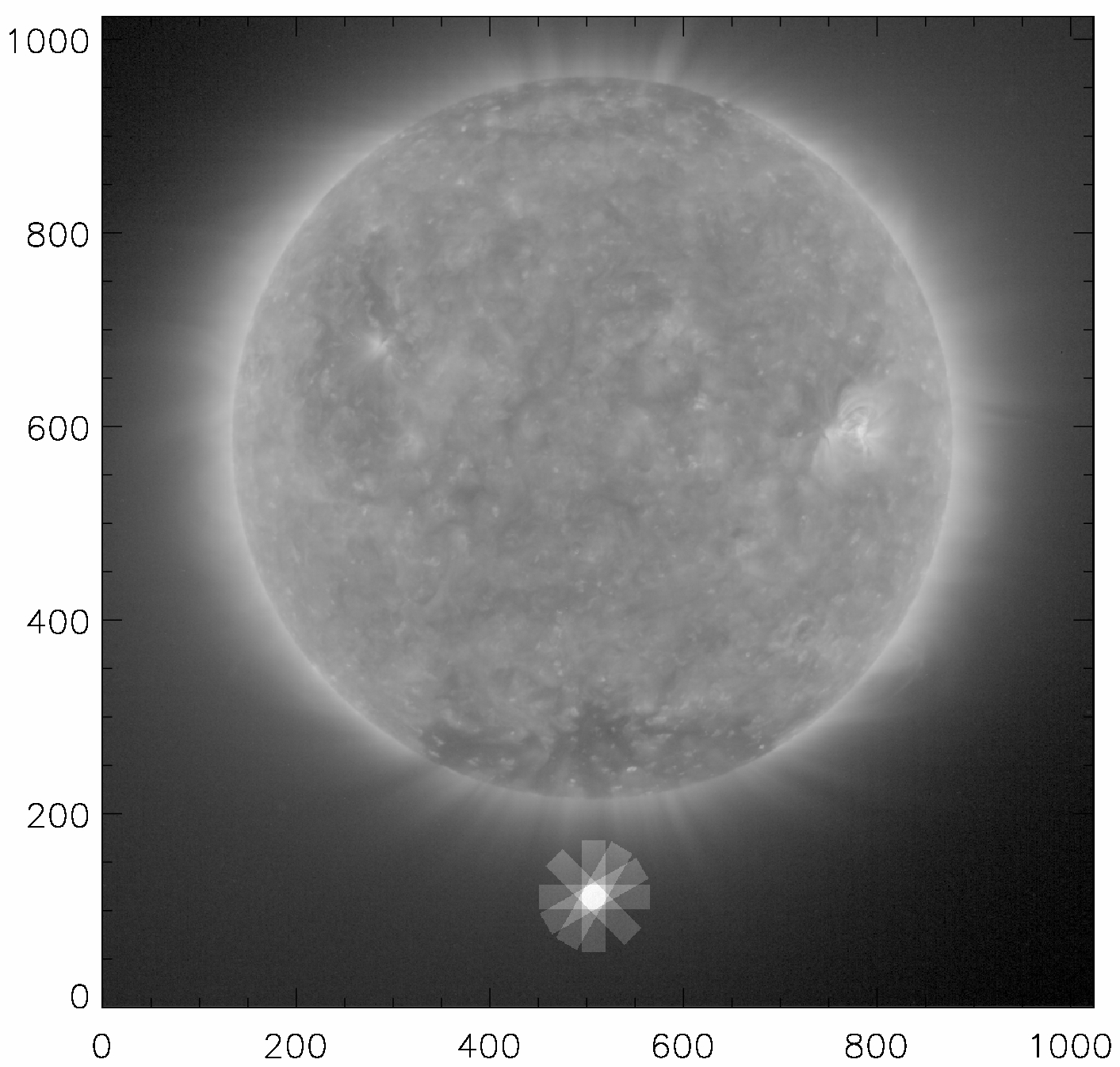} 
\includegraphics[width=.5\textwidth]{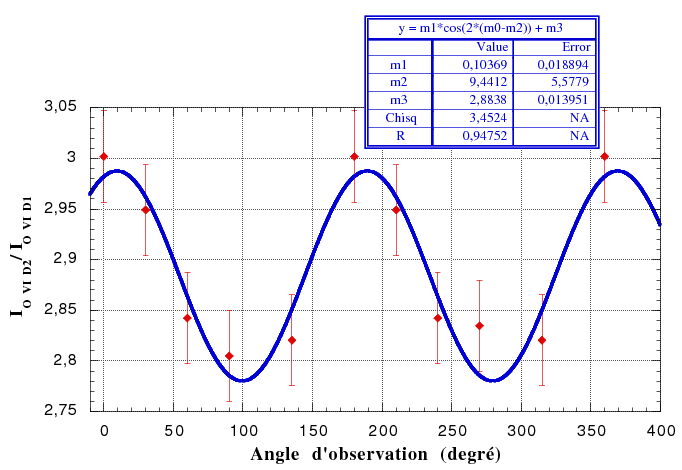} 
\caption{SOHO/EIT 195~{\AA} image of the solar corona on March 19, 1999, with a sketch of the raster sequences performed by SOHO/SUMER in the southern polar coronal hole at $\sim1.3~R_{\odot}$ from Sun center. After \cite{1999A&A...345..999R}.\label{Figure1}}
\end{center}
\vspace{-0.5cm}
\end{figure}

\cite{1999A&A...345..999R,2002A&A...396.1019R} successfully measured the linear polarization of the \ion{O}{vi} 103.2~nm line and interpreted it in terms of the Hanle effect of the coronal magnetic field (see Figure~\ref{Figure1}). This was achieved through observations from SUMER/SOHO. SUMER is sensitive to the linear polarization of the observed light. The observations were made during the roll of SOHO on March 19, 1996, in the south coronal polar hole at 1.3 R$_{\sun}$. For more details on the observation see \cite{1999A&A...345..999R}. \cite{2002A&A...396.1019R} developed models to simulate the observational results taking into account the coronal Hanle effect. They obtained constraints on the coronal field in the polar hole ($B \sim 3$ Gauss). The main result from this work is a clear evidence for the Hanle effect in the strong \ion{O}{vi} 103.2~nm coronal lines. That opens another window for inferring the coronal magnetic field by using different FUV lines with different sensitivities to the magnetic field at different coronal altitudes.

\cite{2009ASPC..405..423M} discuss the possibility of mapping the magnetic fields of coronal loops thanks to a mechanism that produces scattering polarization in permitted lines at EUV wavelengths (e.g., the \ion{Fe}{x} line at 17.4 nm), at which the underlying quiet solar disk is seen dark.

\subsection{Polarization of Coronal Radio Emissions}

In the micro-wavelength range, there are a number of approaches to constrain coronal magnetic fields. The following is a brief description of these methods and examples of the measurements that can be achieved. For more details, we refer the reader to review papers by \citet[][gyroresonance]{White2004}, \citet[][bremsstrahlung]{Gelfreikh2004}, and \citet[][Faraday rotation]{2007A&AT...26..441B}.

\subsubsection{Gyroresonance and Bremsstruhlung Emissions}

Gyroresonance and bremsstruhlung are the two main emission mechanisms of strong circularly polarized radio emissions above solar ARs.

Gyroresonance emissions result from the acceleration of electrons gyrating around magnetic field lines. The resonant interaction of the e- (extraordinary) and o- (ordinary) wave modes is much stronger for the former. The difference in the resonant interaction of the two modes with the electrons gives rise to a strong circularly polarized emission that depends on the strength of the magnetic field.

\begin{figure}[!th]
\begin{center}
\includegraphics[width=.45\textwidth]{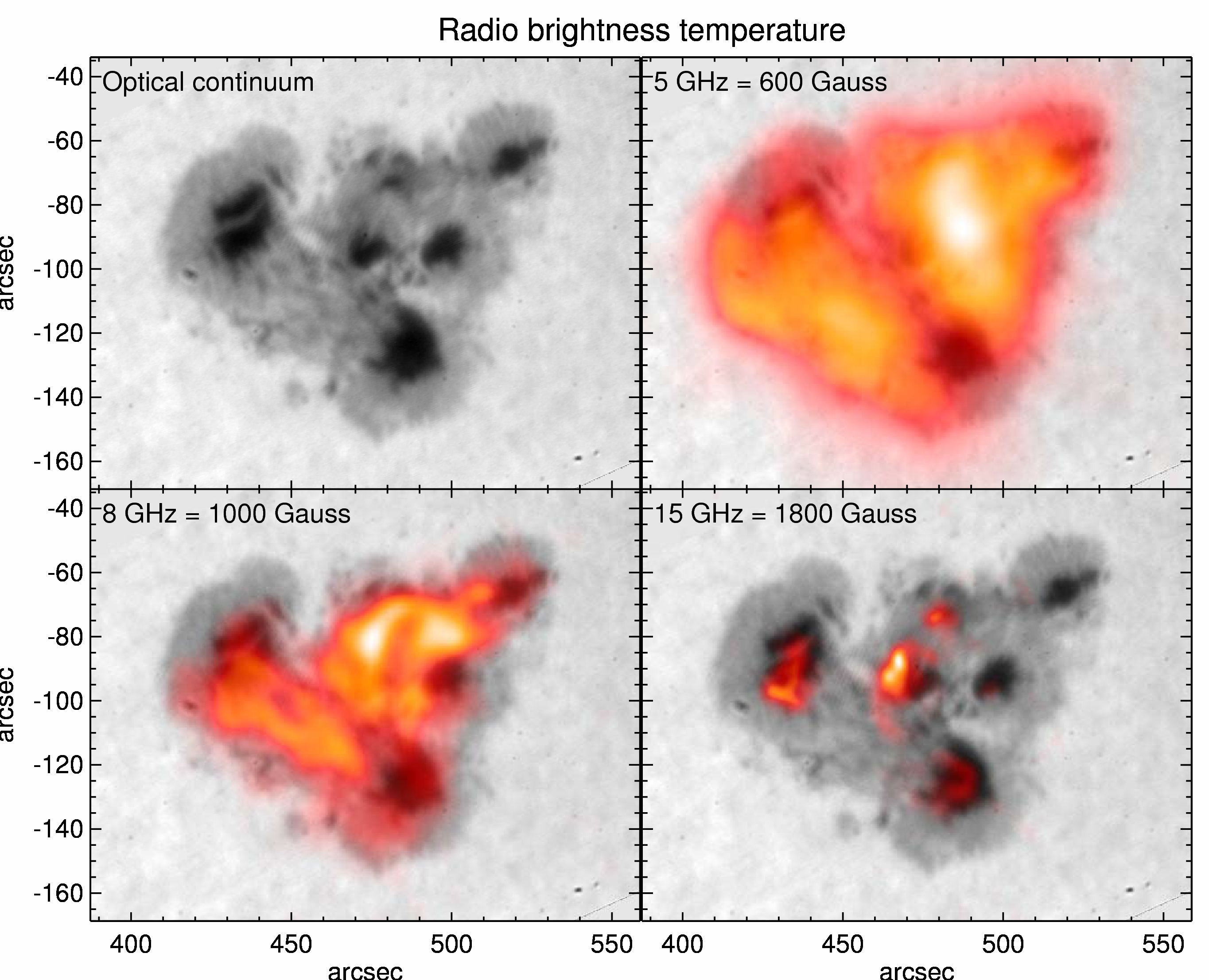}
\includegraphics[height=.45\textwidth,angle=90]{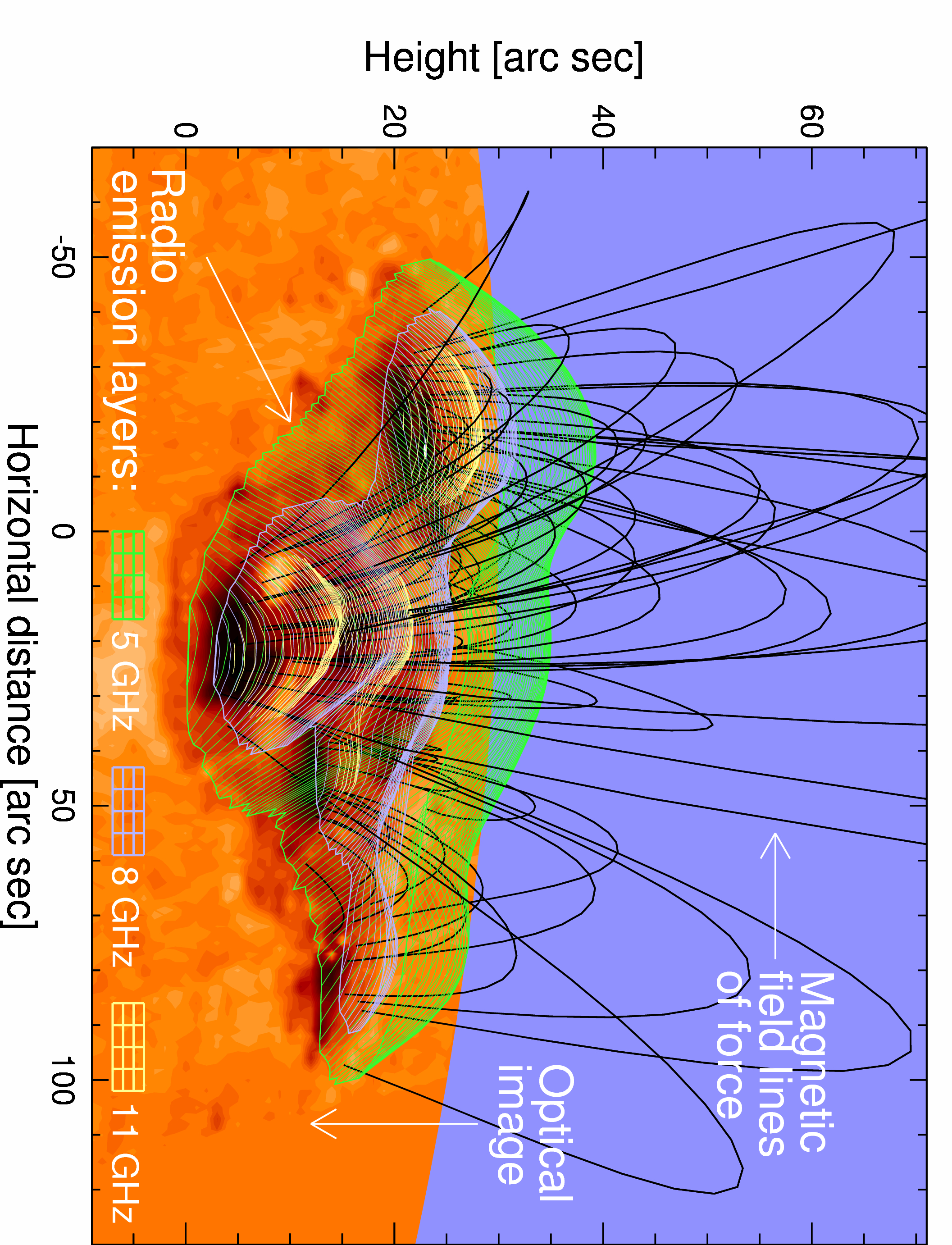}
\caption{Left: Radio brightness distribution of AR6615 (isogauss surfaces) at 5, 8, and 15 GHz from VLA observations superposed on the white light continuum image (top-left). \citep[After][see also {\it{http://www.fasr.org/FASR\_Overview.pdf}}]{1998SoPh..180..193L}. Right: A perspective view of a complex sunspot group (AR6615 on 1991 May 7) in optical continuum is shown with magnetic field lines from a nonlinear force-free calculation and the calculated gyroresonant surfaces in the corona at the radio frequencies: 5, 8, and 11 GHz (B = 600, 950, and 1300 Gauss, respectively). From \cite{2007SSRv..133...73L}.
\label{Gyroresonace_Radio}}
\end{center}
\vspace{-0.75cm}
\end{figure}

The opacity at microwave frequencies is also dominated by the gyroresonance mechanism and is only significant at the few first harmonics (typically $s = 1, 2, 3$) of the electron gyrofrequency \citep[see][]{1968SvA....12..245Z,1997SoPh..174...31W}. This produces narrow resonances that correspond physically to very thin coronal layers of a few tens of kilometers. In addition, the two electromagnetic modes have different optical opacities. Thus, observing the two circular polarizations at the harmonic frequencies allows one to determine isogauss surfaces above ARs. These isogauss surfaces correspond to different coronal heights as illustrated by Figure~\ref{Gyroresonace_Radio} \citep[see][]{1997SoPh..174..175L,1998ApJ...501..853L}. Gyroresonance observations do not provide information on the height of the emission.

Bremsstrahlung emissions are produced by electron collisions with protons and $\alpha$ particles. The coronal magnetic field breaks the degeneracy of the two-wave e- and o-modes of the emitted radiation. The degree of circular polarization is proportional to the LOS-component of the magnetic field (for both optically thin and thick cases). This provides a diagnostic of the average LOS-magnetic fields \citep[see][]{1997ApJ...488..488B}. 

Gyroresonance and Bremsstrahlung diagnostics of coronal magnetic fields complement one another. Gyroresonance diagnostics cover frequencies ranging from $\sim3 - 20$~GHz, which correspond to magnetic field strengths ranging typically from a few 100 Gauss to more than 2.5 KGauss. Bremsstrahlung covers frequencies below 3 GHz and magnetic field strengths from a few Gauss up to a few 100 Gauss.

\subsubsection{Faraday rotation}

The Faraday rotation is the effect of a local magnetic field on the direction of the plane of linear polarization of incident radiation from a background source (galactic or extragalactic or interplanetary space probes such as Pioneer and Helios). It is the combined product of the coronal electron density and the LOS-component of the magnetic field. The refractive index of the coronal medium is different for two circularly polarized components corresponding to the linear polarization of the incident radiation:
$
\displaystyle n^2\approx 1-(\omega_p^2/\omega^2)\left(1\mp(\Omega_e/\omega)\right) ,
$
where $\omega$, and $\omega_p$, $\Omega_e$ are the radiation, plasma, and electron gyro- frequencies, respectively. The propagation speeds of the left- and right-handed circular polarization are given by 
$
\displaystyle v_p = c\mp c\;\Omega_e\;\omega_p^3/(2\;\omega^3).
$
The differential speed results in a rotation of the plane of linear polarization of the emergent signal that is given by
$
\displaystyle \Delta\chi\propto\int_{LOS}N_e{\bf{B}}\cdot{\bf{ds}}.
$

\begin{figure}[!h]
   \begin{center}
\parbox{5.cm}{\includegraphics[width=5.cm]{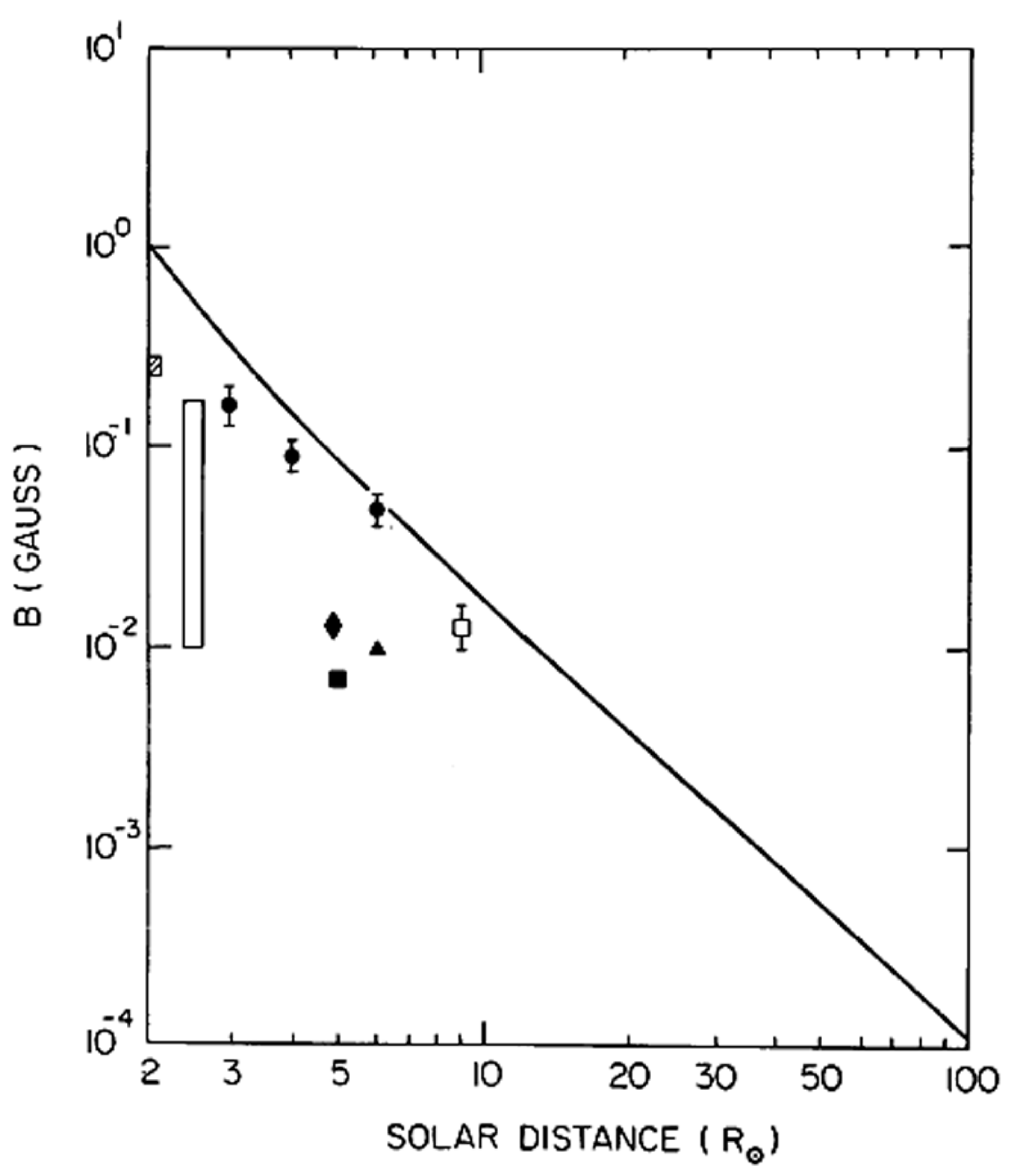}}
\parbox{6.5cm}{\caption{Magnetic field strengths obtained from Faraday rotation measurements. Measurements from other methods (filled circles: gyrosynchrotron; rectangular box: potential field extrapolation; solid line: empirical model) are plotted for comparison. From \cite{1994ApJ...434..773S}. \label{fig9_Radio} }}
   \end{center}
\vspace{-0.5cm}
\end{figure}

Figure~\ref{fig9_Radio} displays averages of the LOS magnetic field strengths obtained from Faraday rotation from natural source at different times and different heights. The same Figure shows how Faraday rotation is reliable compared to other methods \citep[see][]{1994ApJ...434..773S}. The Faraday rotation technique has also been used to infer magnetic fields in transient events and CMEs \citep[see][]{1985SoPh...98..341B}. \cite{1982JGR....87....1H} used Helios data for Faraday rotation measurements and also to study the fluctuating component of the plasma \citep[the magnetic field, electron density, waves, etc.; see also ][]{1999ApJ...525..195M}.

The Faraday rotation method does not, however, allow two-dimensional mapping of the field due to the limited extension of the radio sources and to their transit path in the coronal background. It is also model dependent or needs additional measurements of the plasma parameters such as electron density to determine the field and vice-versa.

\section{Conclusions and Prospects of Coronal Polarization}

Our understanding of complex coronal phenomena, such as the solar activity, coronal heating and the acceleration of the solar wind, depend fundamentally on the measurement of key plasma parameters (e.g., magnetic fields, densities, temperatures, velocities, etc.). Although, photospheric (and to less degree chromospheric) studies are well advanced, it is unlikely that they would help resolving long-standing and challenging coronal phenomena. Dedicated coronal studies are therefore a prerequisite for any advances in understanding coronal manifestations, in particular, the solar magnetic activity and the consequent space weather.

The polarization of coronal emissions at different wavelength regime is an excellent tool to diagnose plasma quantities. Advances made in the coronal polarization are encouraging, in particular in the microwave domain. Achievements made in the optical and FUV-EUV wavelength ranges are, however, modest. This is mainly due to the lack of dedicated instrumentations. 

Future projects, such as the Advanced Technology Solar Telescope (ATST) and the Frequency Agile Solar Radiotelescope (FASR) should allow for much significant advances to diagnose coronal magnetic field through Hanle and Zeeman effects and radio emissions, respectively. The FUV-EUV wavelength ranges provide a number of spectral lines that would allow studying coronal magnetic fields. This cannot, however, be achieved
without instruments designed exclusively for this purpose.

\acknowledgements 
N.-E. Raouafi's work is  partially supported by NASA award number NNX08AJ10G.

%\bibliography{aspauthor}

\end{document}